# Optical properties of oxygen-containing yttrium hydride thin films during and after the deposition

M. Zubkins, I. Aulika, E. Strods, V. Vibornijs, L. Bikse, A. Sarakovskis, G. Chikvaidze, J. Gabrusenoks, H. Arslan, J. Purans

Institute of Solid State Physics, University of Latvia, Kengaraga 8, LV-1063, Riga, Latvia

**Abstract**

The synthesis of the photochromic YHO films is based on the oxidation of deposited yttrium hydride in ambient conditions. The actual state of the films during the deposition process, which is influenced by the deposition pressure and the oxidation caused by the residual gases, is not completely known. We report on the $YH_xO_y$ thin films deposited by reactive pulsed-DC magnetron sputtering. Since the visible light transmittance is closely related to the phase and chemical composition of the films, *in-situ* transmittance measurements during and after deposition are performed. *Ex-situ* spectroscopic ellipsometry is used to determine the optical constants of $YH_xO_y$ throughout the film thickness. In order to obtain metallic $YH_{2-x}$ films, the densest possible structure with a high deposition rate is required, otherwise the films could already be partially transparent during the deposition. The transmittance is higher if deposition pressure is increased. This is because of the oxidation promoted by more porous growth of the microstructure that is observed at the surface and cross-section images of the films. The films exhibit a refractive index gradient perpendicular to the substrate surface, which is related to the porosity and variation of the chemical composition.

Keywords:

photochromism; yttrium oxyhydride; rare earth compounds; thin films; reactive magnetron sputtering; spectroscopic ellipsometry

**Introduction**

Thin films of oxygen-containing rare-earth hydrides (RE: Y, Gd, Dy, Er [1,2]) have been shown to be a promising smart material with a reversible photochromic effect induced by UV-Vis light. They could be used in the next generation energy-efficient windows to provide dynamic and situational solar-controlled properties. In general, due to the higher physico-chemical stability of inorganic compounds over organic compounds, they are more suitable for smart window and sensor applications. The photochromic activity of oxygen-containing RE hydrides strongly depends on the composition [3,4], thickness [5], and illumination conditions [6,7] of the film. For example, oxygen-containing Y hydride films with a proper O/H atomic ratio (~ 0.5–1.0 [8]) show a colour-neutral reversible photochromic effect (contrast up to ≈50%) and a resistivity decrease (from $10^5$ to $10^3$ Ωcm [1]) when illuminated by a solar simulator at room temperature and ambient pressure. The photo-darkened films kept in the dark bleach to their initial clear state in several hours to days.

Despite the extensive research on these compounds in recent years, neither the mechanism responsible for the observed photochromism nor the exact phase of oxygen-containing RE hydrides have been completely understood. Based on the charge neutrality principle, the single phase of rare-earth oxy-hydride (REH$_{3-2x}$O$_x$, $0.5 \leq x \leq 1.5$) is proposed for the photochromic films, where $O^{2-}$ or $H^-$ anions occupy common lattice sites in a face-centred cubic structure [9]. Mixed anion compounds such as oxy-hydrides are an interesting class of materials with a huge potential in future solid-state devices [10]. Multiple ions provide ample opportunities to functionalise these materials. A theoretical study based on the modelling structural transformations caused by anion exchange [11] has predicted a large number of stable chemical compositions and lattice geometries for yttrium oxy-hydride. The cubic $F\text{–}43m$ phase is shown to be the most stable for the chemical formula YHO. In addition, oxygen-rich/hydrogen-poor ternary compositions lead to the formation of thermodynamically unstable phases with respect to the decomposition into binary compounds $Y_2O_3$ and $YH_3$.

The muon spin rotation technique [12] and time-resolved X-ray diffraction [13] show a reversible structural rearrangement with a subsequent change in the electronic state during illumination. The electronic work function of 4.76 eV is reduced by 0.2 eV when the transparent $YH_xO_y$ film is darkened [14], implying that the free electron density becomes higher, and the change in conductivity is likely to be of an electronic nature. Several mechanisms for the photochromic activity such as (i) the formation of metallic domains due to $O^{2-}$ or $H^-$ diffusion [15-17], (ii) hydroxide formation, (iii) colour centres, or (iv) dihydrogen formation [18] are discussed in Ref. [12]. In addition, the photochromic performance has been improved in the $YH_xO_y/WO_3$ composite films and is thought to be due to the synergistic effect based on the $H^-$ diffusion [19]. Further research is urgently needed to confirm the proposed mechanisms, as some of them have been partly disproved. For example, photo-darkening has been observed in $YH_xO_y$ at temperatures as low as 5 K at which the anion diffusion is very limited [20], and a DFT study [18] showed that hydroxides in $YH_xO_y$ are unstable at room temperature. Furthermore, the formation of a metallic phase due to the anion diffusion contradicts the proposed single phase of oxy-hydride. Indeed, two separate phases of $Gd_2O_3$ and $GdH_2$ have been shown by spatially resolved composition analysis at the nanometre scale for the photochromic $Gd_{0.32}H_{0.38}O_{0.31}$ thin films [21]. The measured compressive stress of 5.9 GPa in these films is close to the value at which the photochromic properties are observed for the pure yttrium hydride [22]. The photochromic activity in the $Gd_{0.32}H_{0.38}O_{0.31}$ films is explained by the photon-induced hydrogen diffusion between the two phases.

The photochromic $YH_xO_y$ films can be prepared by the simple exposure of $β\text{-}YH_{2-x}$ films deposited by reactive magnetron sputtering to ambient air atmosphere. During this process, the films oxidise to a certain level due to the high reactivity towards oxygen, and become transparent and photochromic [23]. The oxidation process is accompanied by the release of hydrogen [24] and a change in the electronic state. From a film deposition technology point of view, this raises questions about the influence of residual gases present in the vacuum chamber or other deposition parameters on the composition of the film before the exposure to air as well as the chemical homogeneity throughout the film thickness. The sputtering pressure serves as an indirect control parameter of the composition, and a high photochromic contrast can be achieved over a relatively narrow pressure range [25], although the range boundaries depend on the setup of each individual deposition system and other deposition parameters. In general, the pressure strongly influences the microstructure of films and the grain packing-density according to the Thornton structure zone model [26]. However, in the case of $YH_xO_y$ films, the effect of pressure on the microstructure is not directly demonstrated. The incorporation of oxygen into the films expands the host crystal lattice and opens the optical band gap in the range of 2.8–3.7 eV [23,25].

In this paper we use *in-situ* transmittance measurements to study the state of the films during the deposition by reactive magnetron sputtering and the changes in optical properties after the process in the oxygen containing atmosphere. More information about the deposition process and its influence on the performance of films helps to master the synthesis of photochromic $YH_xO_y$. Magnetron sputtering is among the most widely used types of deposition by the glazing industry because it can be scaled up to large area substrates together with a high growth rate, which is highly important for the large-scale production of smart windows. The synthesis at room temperature also enables deposition on flexible substrates and *roll-to-roll* production. Spectroscopic ellipsometry (SE) is used to determine the optical constants throughout the $YH_xO_y$ films' thickness as a function of the deposition pressure. X-ray diffraction (XRD), electron microscopy, Fourier transform infrared spectroscopy (FTIR) and X-ray photoelectron spectroscopy (XPS) are also used to study the films. This paper is organised as follows: experimental details, oxidation dynamics, kinetics of photo-darkening and bleaching, structural analysis using XRD and electron microscopy, optical properties - spectroscopic ellipsometry (SE), depth profiling by XPS, and conclusions.

**Experimental details**

The film deposition was performed using a vacuum PVD coater G500M (Sidrabe Vacuum, Ltd.). The films were deposited on soda-lime glass, Ti, and Si substrates by reactive pulsed-DC magnetron sputtering from a Y (purity 99.9 %) target in an Ar (6N) and $H_2$ (5N) atmosphere ($H_2$ to Ar gas flow ratio of 1:2). The $YH_xO_y$ phase was obtained by the intentional introduction of oxygen (5N) gas into the chamber up to ≈ 400 Pa in 30 min (dosage ≈ $10^9$ Langmuir) after the deposition. The capping layer (except for the FTIR measurements) was not used to protect the films from further oxidation; however, the samples were stored in an Ar containing atmosphere for most of the time. The characterisation was performed within a few hours to a few days after the synthesis of the $YH_xO_y$ samples. The exception was some measurements in which the stability of the films was tested on purpose. The samples for the FTIR measurements were deposited on Si substrates. Some of them were covered with approximately 300 nm Al film without breaking the vacuum.

The substrates were ultrasonically cleaned with acetone, 2-isopropanol for 15 min each, rinsed with distilled water, and then dried under a flow of $N_2$ gas before deposition. A planar balanced magnetron with target dimensions 150 mm × 75 mm × 3 mm was used. The magnetron was placed against the grounded substrate holder at a distance of 10 cm (the substrate facing the target axis). The substrates were not heated intentionally during the deposition. Before the deposition process, the chamber (≈ 0.1 m$^3$) was pumped down to $6 \times 10^{-4}$ Pa by a turbo-molecular pump backed with a rotary pump. The target was sputtered at a constant average power of 200 W. The pulse frequency was set at 80 kHz with the reverse time of 2.5 µs. The deposition was performed for 20 min at different sputtering pressures from 0.40 to 2.65 Pa. The pumping speed was altered by a throttle valve to set the necessary sputtering pressure. The thickness of the films was in the range from 330 to 460 nm.

The *in-situ* transmittance of the films during the deposition and oxidation process was measured using CMOS detector Avantes StarLine AvaSpec-ULS2048CL-EVO, and a tungsten-halogen light source Avantes AvaLight-HAL-S-MINI. The *ex-situ* photochromic response was tested using UVA light from a 15 W (≈2.4 mW/cm$^2$) lamp under ambient conditions (Fig. S1 in supplementary material). The energy of the source light at the maximum intensity is 3.3 eV with the spectral width of 0.13 eV (the FWHM of the peak). A tungsten-halogen lamp with the power density of ≈3.3 mW/cm$^2$ was used for the probing. No photo-darkening due to the probing light was observed.

The thickness of the films was determined by a profilometer, CART Veeco Dektak 150. The structure of the samples was examined by an X-ray diffractometer with Cu-Kα radiation, Rigaku MiniFlex 600. The film surfaces were characterised by scanning electron microscopy (SEM). The lamellae were prepared by focused ion beam (FIB) and visualised by scanning transmission electron microscopy (STEM) using a Thermo Scientific Helios 5 UX dual-beam microscope. Before lamellae preparation, the surface of the samples was sputtered with a 30 nm thick Au layer and a 1.5 µm thick Pt layer to protect the surface from FIB exposure.

*Ex-situ* transmittance spectra, optical properties and film thicknesses were obtained by means of spectroscopic ellipsometer (SE) WOOLLAM RC2 in the spectral range from 210 to 1690 nm or from 5.9 to 0.7 eV. The main ellipsometric angles Ψ and Δ were measured at incident angles from (45-75)° with a 5° step. The complex dielectric function dispersion curves (refractive index *n* and extinction coefficient *k*) were modelled with Drude oscillator (DO) and 3-4 Lorentz or Gaussian oscillators (GO) for conductive thin films (absorbing in the whole spectral range), and with one Tauc-Lorentz (TLO) and one GO for semiconducting thin films (transparent in the VIS region up to around 3 eV). SE modelling was performed by CompleteEASE software.

The optical constants of surface roughness were derived by mixing the optical constants of the underlying materials with the optical constants of voids ($n = 1$, $k = 0$). The Bruggeman's model – Effective Medium Approximation (EMA) is used to calculate the optical constants of this mixed layer, assuming 50% void content. While the effective roughness layer approach is certainly an approximation to the actual samples, this approach works extremely well for modelling SE data when the size of the surface is much less than the wavelength of light used to measure the samples.

Optical gradient (*n* and *k* variation within the depth of the film) was modelled by applying the parameterised gradient (variation of the oscillator fitting parameters with the film thickness) and a simple graded model with percentual variation of *n*, *k* (inhomogeneity) considering the films as a stack of 11-13 slices (lowest MSE observed). The porosity of the films was obtained by modelling the film as graded EMA - percentual mixture of material and voids optical properties and their variation with the film thickness. The MSE (mean square error) values for the conducting films were between 2 and 5. The MSE for other films were higher, ranging from 5 to 12. The inhomogeneity models showed high efficiency, reducing the MSE on average from around 50-60 to 8. The thickness and porosity of some films were evaluated by SEM images analysis with ImageJ software.

FTIR absorbance spectra were measured by a VERTEX 80v vacuum FTIR spectrometer. The experiments were performed in the range from 1000 to 4000 cm$^{-1}$, with the interferometer working in vacuum and with a resolution of 4 cm$^{-1}$. Uncoated Si was used as a background. The non-capped films were measured in transmittance mode and the Al capped films in reflectance mode by the irradiation through the substrate. Details on X-ray photoelectron spectroscopy (XPS) analysis are described in [27]. Depth profiling of the films was accomplished using the Ar$^+$ ion gun operated at 2 kV and rastered over a 2 × 2 mm 10 seconds for each layer. Since XPS cannot detect hydrogen, the composition analysis is given as the concentration ratio between oxygen and yttrium.

## 3. Results and discussion

### 3.1. In-situ transmittance measurements

Since the transmittance of YH$_x$O$_y$ films is closely related to the O/H atomic ratio, it was measured *in-situ* during the deposition and the post-oxidation. These measurements show the influence of sputtering pressure (including hydrogen partial pressure) and residual gases on the state of the films and the optical properties. In the deposition pressure range of 0.40 to 2.65 Pa, the hydrogen partial pressure varied from 0.07 to 0.45 Pa, respectively. The residual gas composition at the base pressure of $6 \times 10^{-4}$ Pa can be found in Fig. S2 (supplementary material), which consists of 82% ($4.7 \times 10^{-4}$ Pa) $H_2O$, 14% ($7.4 \times 10^{-5}$ Pa) $N_2$, 2% ($1.3 \times 10^{-5}$ Pa) $O_2$, and other gases < 2%.

The transmittance spectra of the films measured immediately after the deposition are shown in Fig. 1(a). One can see that the films become partly transparent at the deposition pressure greater than approximately 1 Pa (or 0.15 Pa hydrogen partial pressure), despite the fact that the films become thicker with the deposition pressure (see Table 1 in Section 3.3. and Fig. S3 in supplementary material). The formation of transparent $\gamma$-$YH_3$ is unlikely due to the low pressure conditions. Although the ($\beta$-$YH_2$)–($\gamma$-$YH_3$) coexisting phase plateau at the pressure-composition isotherm (at room temperature) is observed at the hydrogen partial pressure of approximately 0.1 Pa [28,29], the transparency appears only in the $\gamma$-$YH_3$ rich composition at significiently higher pressures [30]. In addition, the $\gamma$-$YH_3$ phase cannot be detected *ex-situ* for the non-capped films because it is unstable and dissociates at room temperature into $YH_2$ and $H_2$ [31,32].

The main reason for the increase in transmittance at deposition pressures above 1 Pa should be the oxidation at given residual gas ($H_2O$ and $O_2$) levels due to the more porous structure of the films. Yttrium is able to dissociate water or oxygen to form stable yttrium oxide ($Y_2O_3$). The transparent $Y_2O_3$ films can be prepared by the reactive magnetron sputtering using oxygen partial pressure as low as $1 \times 10^{-3}$ Pa [33]. Yttrium hydride is highly reactive towards oxygen due to the low electronegativity of yttrium ($\chi_Y$ = 1.22). In addition, oxygen has a higher electronegativity ($\chi_O$ = 3.44) than hydrogen ($\chi_H$ = 2.20). As a result, oxygen can easily penetrate the yttrium hydride structure and replace significant amounts of hydrogen.

The composition of our samples has not been directly measured, but comparing the transmittance spectra with other studies [2,15,34,35], we can identify an $YH_{2-\delta}$ phase. This is evidenced by the characteristic transmittance peak around 700 nm [34], which can be seen in the enlarged spectra of the films synthesised below 1 Pa (Fig. S4 in the supplementary material). After air exposure, a certain level of oxygen contamination can be expected for these films, as they are not protected by a capping layer.

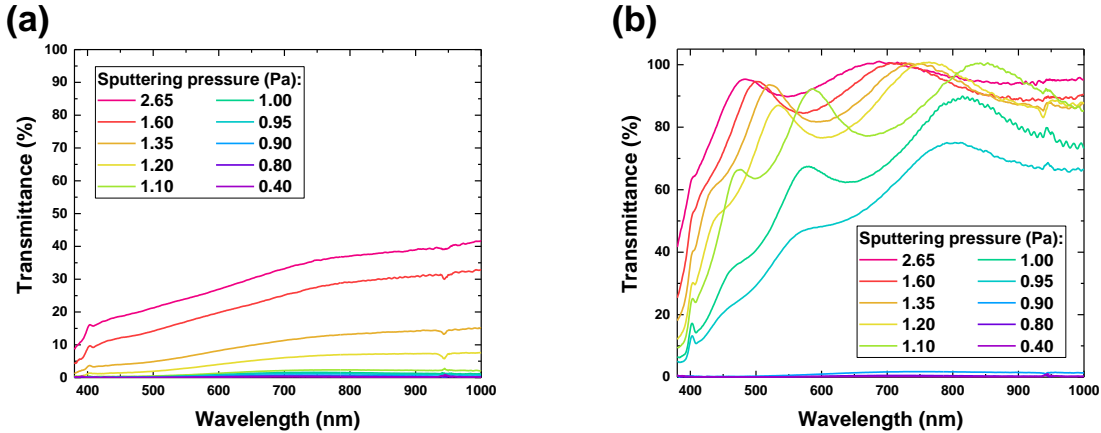

**Fig. 1.** *In-situ* transmittance spectra of the deposited $YH_xO_y$ films on soda-lime glass immediately after the deposition (a) and 30 min after introduction of oxygen ($Q(O_2)$ = 10 sccm with stopped pumping) into the chamber up to 400 Pa (b).

After the deposition, pumping was stopped and at the same time the continuous oxygen (5N) gas flow of 10 sccm was set into the chamber to gradually oxidise the films. The transmittance spectra after 30 min of oxidation ($\approx 10^9$ Langmuir) are shown in Fig. 1(b). All the films deposited at the pressure $\geq$ 0.95 Pa become highly or partly transparent. Again, the average transmittance in the measured range is higher for the films deposited at higher pressure. The partly transparent films due to the incomplete oxidation, deposited at 1.00 and 0.95 Pa, exhibit a photochromic effect (Section 3.2.). These films appear yellowish due to the absorption of light in the blue and green parts of the spectrum.

In some cases, the films produced at around 0.90–0.95 Pa exhibit both transparent and dark areas when removed from the chamber. This is due to lateral variation of density. The dark areas slowly decrease over time. The newly opened (transparent) areas show a high photochromic activity. The films deposited at lower pressures (< 0.90 Pa) do not oxidise to the level to become transparent, however, a minor increase in the transmittance around 700 nm is detectable for the films deposited at 0.90 and 0.80 Pa. These metallic films do not turn photochromic with time either. We can conclude that in order to produce the purest possible yttrium hydride film, the lowest possible pressure with a high deposition rate should be used. The *ex-situ* transmittance spectra can be seen in Fig. S5 in the supplementary material. The films show a sharp absorption edge between 300 and 400 nm depending on the deposition pressure.

The transmittance at several wavelengths was measured over time to evaluate the rate of change. An example at 420 nm (other wavelengths can be seen in Fig. S6) is shown in Fig. 2(a). Although the curves resemble exponential behaviour, they cannot be fitted by one exponential function. For this reason, the time constant was defined as the time required for the transmittance to reach half of its final value at $t$ = 30 min. We aware that transmittance would continue to increase slowly even after 30 min., and longer data collection times would slightly increase time constants. It is claimed in Ref. [3] that a metastable state is reached after a few days in air. Time constants in our case reflect the rapidness of the process, which is closely related to the oxidation. The oxidation occurs more rapidly when the deposition pressure is increased. This is clearly seen in Fig. 2(b) where the time constants are plotted against several wavelengths and deposition pressures. For example, the time constant of 187 s at 420 nm and 0.9 Pa reduces down to 9 s at 420 nm and 2.65 Pa. In addition, there is also a tendency for the time constant to reduce at higher wavelengths. The time constant of 29 s at 420 nm decreases to 7 s at 980 nm (both for 1.35 Pa). Fig. 2(b) does not contain the time constants at 0.40 and 0.80 Pa because the change in transmittance with time is at the noise level, i.e., it does not change.

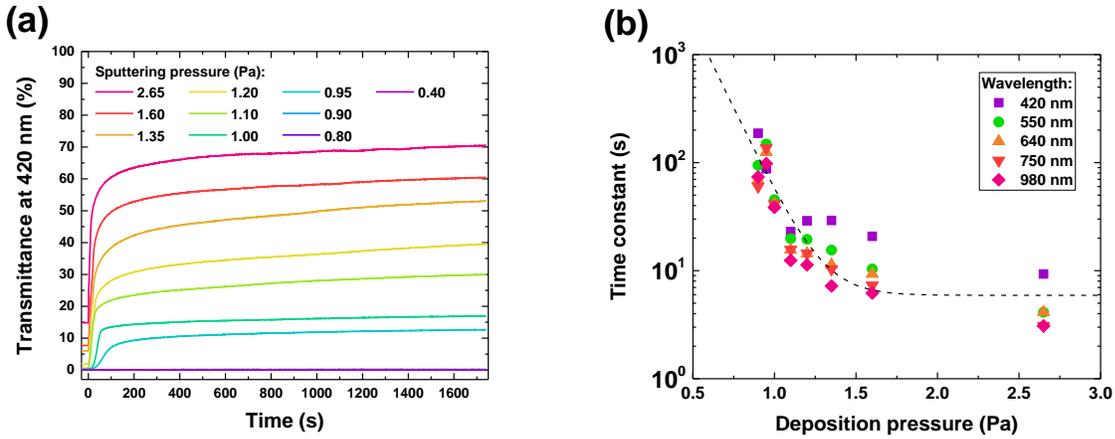

**Fig. 2.** Transmittance at 420 nm over time for different deposition pressures (a). The oxygen introduction is set to $t = 0$. The time constant (defined as the time required for the transmittance to reach half of its final value at $t = 30$ min) at several wavelengths as a function of deposition pressure (b).

The composition shortly after the exposure to air ranges approximately from $YH_{1.25}O_{0.90}$ to $YH_{0.55}O_{1.25}$ according to the *ex-situ* SE analysis and literature, see Section 3.4. We conclude that high oxygen uptake occurs within a few minutes. The gradual change in the composition over several months is shown in Ref. [24], which indicates that the naturally formed oxide layer on the surface of the films does not stop further oxidation. The composition before and after controlled oxidation has been measured for e-beam grown YHO films [36]. $YH_2$ films with the oxygen content of 6.5 at.% turn photochromic after an oxidation dose of $9 \times 10^6$ Langmuir (at least two orders of magnitude lower than used in this study). The photochromic phase is assigned the chemical formula $YH_{1.16}O_{0.84}$, which is close to what we estimated for the photochromic films.

Films grown by e-beam evaporation oxidise and lose their photochromic effect much faster after air exposure than magnetron sputtered films. In general, this is due to a less dense microstructure. From the perspective of our measurements, it must also be taken into account that transmittance is not a linear function of O/H ratio. Although photochromic activity is observed over a wide range of O/H ratio, the photochromic contrast can decrease rapidly depending on the oxygen content [4]. This points out the stability issues and the consequential loss of the photochromic activity if this material is not protected to stabilise the composition.

*3.2. Kinetics of photo-darkening and bleaching*

The photochromic contrast of approximately 10% after 20 min of UVA irradiation is measured for the 750 nm thick $YH_xO_y$ film deposited at 0.95 Pa (Fig. 3(a)). The value is lower compared to the highest contrasts (up to 40%) measured in other studies [15,23,37]. There are several reasons for this. The photochromic effect in YHO has a bulk nature [5]. The high contrast is often shown for thicker films. In addition, the contrast depends also on the intensity and energy of light. On the other hand, the film deposited at 0.95 Pa shows relatively short respond times. It has been shown that the respond times are shorter at higher O/H ratios [3]. Undoubtedly, the photochromic response can be adjusted by the thickness and/or $H_2$/Ar ratio during deposition (Fig. 3(b)). The photochromic contrast is approximately 15% if the thickness is increased up to 1000 nm. It can be further increased to 25% by increasing $H_2$/Ar to 1/1 (even at a smaller film's thickness – 650 nm).

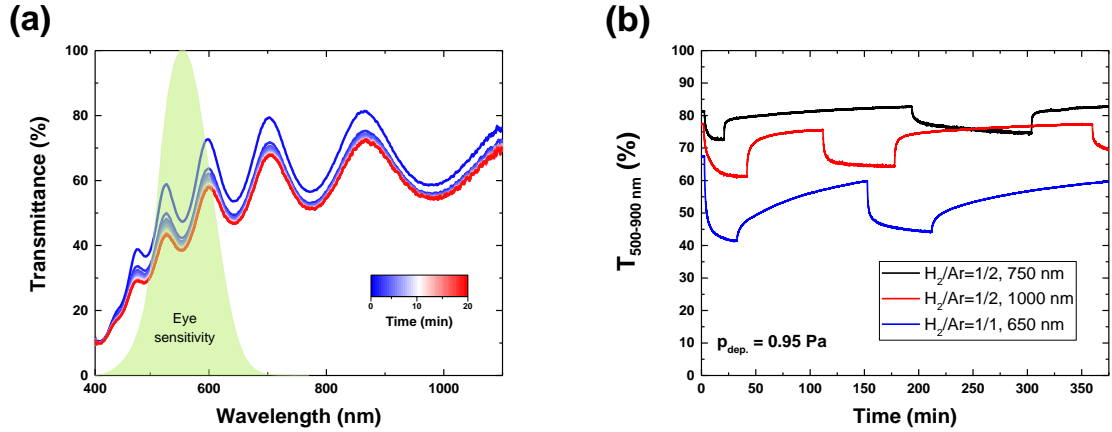

**Fig. 3.** Series of transmittance measurements during the 20 min illumination by UVA light (≈2.4 mW/cm$^2$) of the 700 nm YH$_x$O$_y$ film deposited at 0.95 Pa (a). Average transmittance measured between 500-900 nm during illumination/dark cycling of the films with different thicknesses and sputtering gas flow H$_2$-to-Ar ratios (total pressure of 0.95 Pa was kept constant) (b).

*3.3. Structural analysis using XRD, FTIR and electron microscopy*

The X-ray diffractograms for the YH$_x$O$_y$ films on soda-lime glass were recorded over a range for 2θ of 10°–70° and can be seen in Fig. S7 (in the supplementary material). For more accurate analysis of the structural changes with the deposition pressure, the enlarged X-ray diffractograms between 26° and 40° are shown in Fig. 4(a). Several relatively low intense XRD maximums at 29, 33, 48, 57, and 61° can be observed, which correspond to the cubic lattice planes (111), (200), (220), (311), and (221), respectively. According to the powder diffraction card ICDD 04-002-6939 of *β*-YH$_2$ (*Fm-3m*), the maximums are strongly shifted to lower angles indicating an expansion of the lattice after the incorporation of oxygen, which has already been shown in other studies [13,23,34,38]. The lattice parameter *a* increases from 5.29 to 5.39 Å as the sputtering pressure increases from 0.40 to 1.10 Pa. However, at 2.65 Pa it again reduces to 5.37 Å. The value of parameter *a* for the dark film deposited at 0.40 Pa is already larger than of pure $YH_{1.98}$ ($a = 5.20$) [39], which indicates the oxidation caused by the non-capped films. However, it should be noted that the deposition by the reactive magnetron sputtering is a non-equilibrium process, and some disorder can be expected compared to ideal structures. The obtained values of 5.29 Å at 0.40 Pa and 5.39 Å at 1.10 Pa are in good agreement with the stable YHO lattice geometries (*F*-43*m* and *P*-43*m*) theoretically predicted in Ref. [11]. The average size of the crystallites according to the Scherrer equation is approximately 9 nm for the all samples. In addition, the texture, i.e., the intensity ratio between the (111) and (200) XRD maximums differs significantly for the transparent ($p_d > 0.9$ Pa, $I_{(200)}/I_{(111)} \approx 0.90$) and metallic films ($p_d < 0.9$ Pa, $I_{(200)}/I_{(111)} \approx 0.03$).

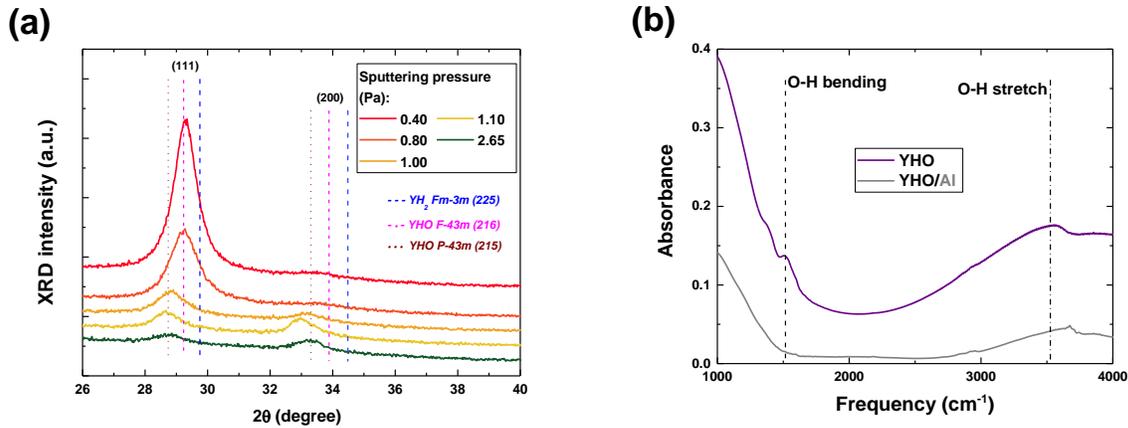

**Fig. 4.** X-ray diffractograms of the $YH_xO_y$ films produced by reactive ($H_2$ to Ar gas flow ratio of 1:2) pulsed-DC magnetron sputtering without intentional substrate heating at different deposition pressures (a). The angles of the Bragg peaks for the cubic *β*-$YH_2$ (ICDD 04-002-6939) and the cubic YHO phases from Ref. [11] are shown by vertical dashed lines. Absorption spectra in the mid-IR region of the $YH_xO_y$ and $YH_xO_y$/Al films deposited at 1.10 Pa on Si substrates (b).

The formation of hydroxides anions would interfere with the growth of the photochromic oxyhydride phase. The presence of O-H species was investigated by FTIR spectroscopy. Fig. 4(b) shows the absorption spectra in the mid-IR region of the films deposited at 1.10 Pa. The spectrum of the non-capped film contains the absorption bands at about 3500 and 1500 cm$^{-1}$, which are attributed the stretching and bending vibrations of O–H bond, respectively. The bands have a relatively low intensity. Most O–H groups form in the films when air is introduced into the chamber because these bands are almost undetectable for the capped film.

The surface morphology and the cross-sectional images of the films obtained by electron microscopy are shown in Fig. 5 and Fig. 6. In the study of the photochromic yttrium hydride films by atomic force microscopy (AFM) [40], no significant changes of the surface morphology were observed with the deposition pressure. In our case, the situation is different. The surface morphology of the film deposited at 0.40 Pa is smooth with a fine-grained structure. At higher deposition pressures (0.90 and 1.00 Pa), the grains grow larger and exhibit an arbitrary shape. In the meantime, the voids between these grains become visible (Fig. 5(c)). The average grain size of approximately 50 nm is slightly less than the 100-200 nm reported in Ref. [33]. By increasing the pressure to 2.65 Pa, the morphology becomes even more porous with an increased number of voids, although the grains again appear smaller.

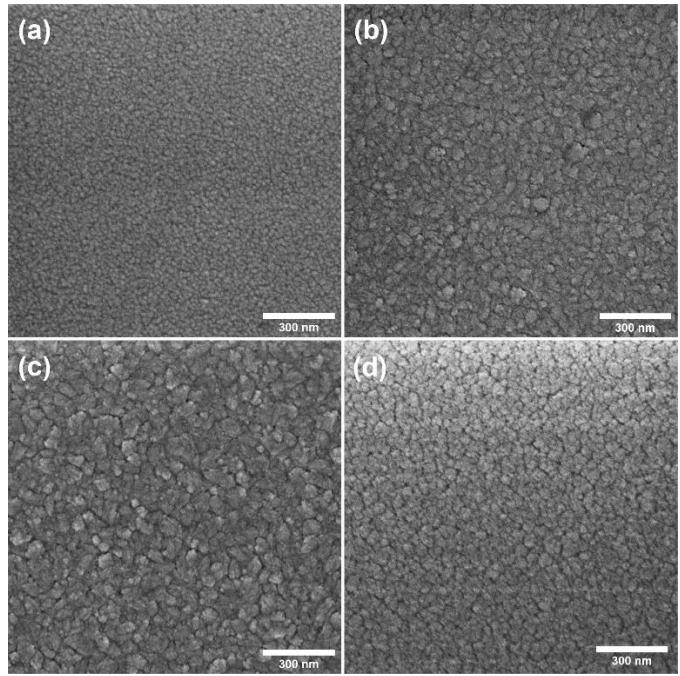

**Fig. 5.** Surface SEM images of the YH$_x$O$_y$ films produced by reactive (H$_2$ to Ar gas flow ratio of 1:2) pulsed-DC magnetron sputtering without intentional substrate heating at different deposition pressures: 0.40 Pa (a), 0.90 Pa (b), 1.00 Pa (c), and 2.65 Pa (d).

The cross-sectional images show the same trend with the increase of the deposition pressure. The films deposited at 0.40 and 0.90 Pa are dense with featureless cross-sectional images (Fig. 6(a,b)). When the pressure is increased to 2.65 Pa, a fine-grained structure is only observed within the initial stage of the YH$_x$O$_y$ film, which evolves into a columnar growth with the vertical voids perpendicular to the substrate surface. The results obtained by the electron microscopy are clearly in line with the Thornton structural zone model [26], as is also in the case of YH$_x$O$_y$ deposition. The film deposited at 2.65 Pa is highly transparent due to its more porous microstructure, which allows oxygen to penetrate the film more easily and promotes the oxidation process. High surface porosity has been predicted in the study of the oxygen-containing yttrium hydride films deposited at 6 Pa by elastic recoil detection analysis [24]. The enhanced presence of oxygen at the grain boundaries for the photochromic $Gd_{0.32}H_{0.38}O_{0.31}$ films has been observed by energy-dispersive X-ray spectroscopy mapping [21].

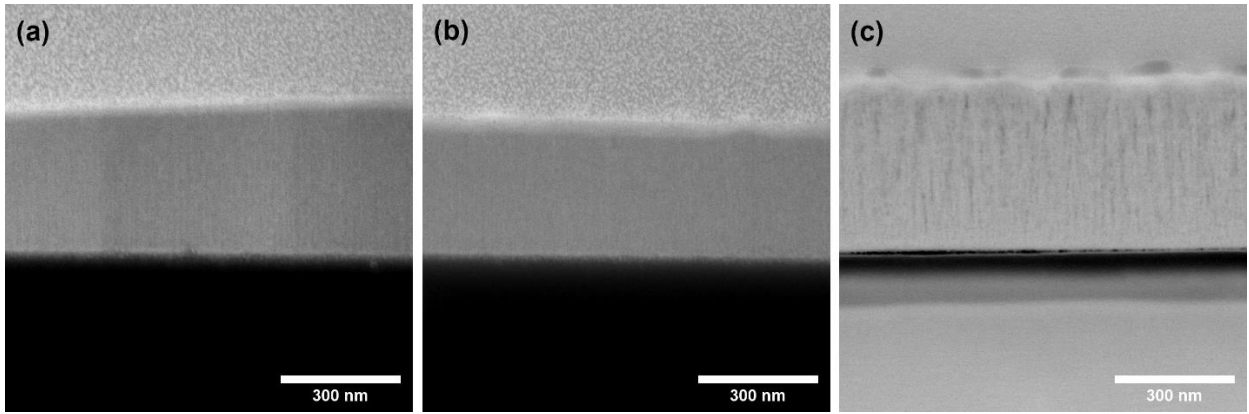

**Fig. 6.** Lamellae cross-section STEM images of the $YH_xO_y$ films produced by reactive ($H_2$ to Ar gas flow ratio of 1:2) pulsed-DC magnetron sputtering without intentional substrate heating at different deposition pressures: 0.40 Pa (a) and 0.90 Pa (b) visualised by high-angle annular dark-field technique, and 2.65 Pa (c) visualised by bright field technique.

*3.4. Optical properties - spectroscopic ellipsometry (SE)*

Fig. 7 summarises optical constants of the $YH_xO_y$ films sputtered at different pressures. The refractive index *n* and extinction coefficient *k* values for the films sputtered at pressures > 0.90 Pa are given as mean values from the top and bottom of the film, since these films present an optical gradient (Fig. 8 and Fig. 9). For these films, *n* and *k* increase in the *E* range from 0.7 eV to ~ 4.5 eV with an optical band gap around 2.5-3.5 eV (Fig. 7(a, b)), which is a distinctive attribute for semiconductor materials. The values of *n* and *k* decrease and the amplitude between maximum and minim values of *n* and *k* also decrease with the increase of sputtering pressure from 0.95 to 2.65 Pa. This is due to the changes in the density of the material and, consequently, in its composition. XRD analysis revealed that with the increase of the sputtering pressure the texture changed and maximums slightly shifted towards lower 2θ values (Fig. 4(a)).

It can be seen that in the spectral region from 0.7 eV up to ~ 4.0 eV, the *n* increases and *k* decreases with the increase of the sputtering pressure from 0.40 Pa to 0.90 Pa (Fig. 7(c,d)). This is due to the oxidation process: the increase of the sputtering pressure causes the formation of semiconducting $YH_xO_y$ thin films; at low pressure ~ 0.40 Pa only a metallic *β*-$YH_{2-x}$ thin film is formed. The films sputtered at pressure < 0.95 Pa show a significant increase in *n* and *k* at $E < 2$ eV (Fig. 7(c,d)), which is a typical characteristic for materials with free charge carriers. Table 1 summarises essential values obtained from SE data modelling and interpretation. The obtain values of *n* are in good agreement with the values found in other studies [16,37]. It can be seen that the electrical resistivity of the films increases with the sputtering pressure: the more the film gets oxidised the less conductive it becomes.

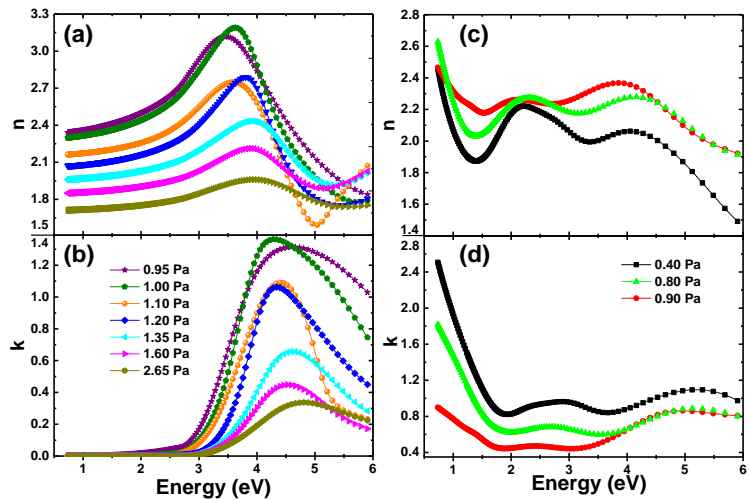

**Fig. 7.** The refractive index $n$ (a, c) and extinction coefficient $k$ (b, d) as functions of photon energy $E$ and sputtering pressure for the $YH_xO_y$ films produced by reactive ($H_2$ to Ar gas flow ratio of 1:2) pulsed-DC magnetron sputtering without intentional substrate heating.

**Table 1.** The roughness $d_R$, resistivity $\rho$, scattering time $t_{sc}$, film thickness $d$ (three sources: profilometer, TEM and SE), Tauc band gap energy $E_g$, mean value of refractive index at 1.960 eV (632.8 nm), and porosity (two sources: SEM and SE) of the $YH_xO_y$ films deposited at different sputtering pressures. The films sputtered at pressure > 0.90 Pa have a non-uniform thickness and an optical gradient. The error bars for $d$ obtained from SE are given based on modelling reflection and transmission data; non-uniform thickness contribution should be considered as well. A thickness non-uniformity of 15% was observed for the films sputtered at 2.65 Pa. A thickness non-uniformity of 2-5% was observed for the rest of the films sputtered in the range from 0.95 to 1.60 Pa. Considering this fact, $d$ values agree when comparing different techniques.

| $p_d$ (Pa) | $d_R \pm 0.1$ (nm) | $\rho \times 10^{-4}$ ($\Omega$cm) | $t_{sc}$ (fs) | $d$ (nm) Profile.; cross-section TEM | $d$ (nm) SE | $E_g$ (eV) | $n$ at 1.96 eV | Porosity (%) cross-section TEM | Porosity (%) SE |
|---|---|---|---|---|---|---|---|---|---|
| 0.40 | - | 6.85 | 0.399 | 422; | 330 ± 1 | - | 2.15 ± 0.03 | 3 - 4 | - |
|  |  |  |  | 375 ± 20 |  |  |  |  |  |
| 0.80 | 7.9 | 13.26 | 0.494 | 406 | 373 ± 5 | - | 2.22 ± 0.04 | | - |
| 0.90 | 6.9 | 31.37 | 0.622 | 402; 341 ± 13 | 369 ± 16 | 0.30 ± 0.30 | 2.25 ± 0.03 | 3 - 4 | |
| 0.95 | 11.6 | All samples in this section are semiconducting or insulating and have thickness non-uniformity between 2-15% and optical gradient. | | | 355 ± 2 | 2.68 ± 0.01 | 2.49 ± 0.06 | | 2.1 - 6.5 |
| 1.00 | 12.3 | | | | 378 ± 1 | 2.66 ± 0.01 | 2.43 ± 0.07 | | 0.9 - 5.2 |
| 1.10 | 14.5 | | | | 423 ± 8 | 2.99 ± 0.03 | 2.25 ± 0.05 | | 0.2 - 3.5 |
| 1.20 | 14.7 | | | | 403 ± 6 | 2.88 ± 0.01 | 2.16 ± 0.03 | | 0.1 - 3.2 |
| 1.35 | 16.1 | | | | 420 ± 2 | 2.75 ± 0.01 | 1.95 ± 0.07 | | 0.1 - 3.0 |
| 1.60 | 17.0 | | | | 435 ± 11 | 2.78 ± 0.01 | 2.14 ± 0.03 | | 0.1 - 3.0 |
| 2.65 | 17.9 | | | 411 ± 20 | 461 ± 3 | 2.79 ± 0.07 | 1.86 ± 0.02 | 40 | 34 - 45 |

Based on the electron microscopy analysis which demonstrated that the films' porosity increases with the increase of the sputtering pressure (Fig. 5, 6), an alternative model was applied to analyses SE data. The introduction of porosity into the model presented the same refractive index tendency:

$n$ decreases for the films sputtered at pressure > 1.00 Pa (Fig. 8). The obtained $n$ and $k$ values are not effective but physical, which can be affected only by structural and compositional characteristics of the material. However, the porosity models showed MSE error from 1.5 to 2 times higher respect to the optical gradient models. The porosity modelling showed that the YH$_x$O$_y$ films sputtered between 0.95 and 1.60 Pa have a porosity in the range of 0.1 - 6.5%; with the gradient destruction the porosity increases in the direction from the substrate to the surface of the films (Table 1). Thus, the SE modelling is also consistent with the Thornton structural zone model [26], where the fine-grained structure near the substrate evolves into the columnar growth with larger voids between grains. The highest porosity is observed for the films sputtered at 2.65 Pa; it varies from 34% up to 45% (from substrate to the surface). This indicates a significant change in the growth of the films when the pressure is increased from 1.60 to 2.65 Pa. It should be noticed that with an increase of sputtering pressure, the films roughness increases (Table 1). The increase in porosity and roughness at higher pressures is related with an increased probability of collisions in the gas phase, where sputtered atoms lose their kinetic energy and thus their mobility on the surface of a growing film. In addition, Table 1 shows that the thickness tends to increase with the deposition pressure.

The $n$ and $k$ values at 400 nm wavelength as a function of sputtering pressure are given in Fig. 8. The $n$ near the substrate ($n$ Bottom) is higher with respect to the $n$ values near the surface ($n$ Top) for the films sputtered at pressure > 0.95 Pa. While $k$ decreases with the increase of sputtering pressure, $n$ reaches its maximum value for the films sputtered at ~ 1.00 Pa. There is no difference in the values of $n$ near the top and near the bottom of the films sputtered at pressure < 0.95 Pa. There is a slight variation of $k$ comparing the corresponding $k$ values for the top and for the bottom of the films sputtered at pressure of (1.20 - 1.35) Pa: $k$ has higher values near the bottom of the films respect to the $k$ near the top of the film.

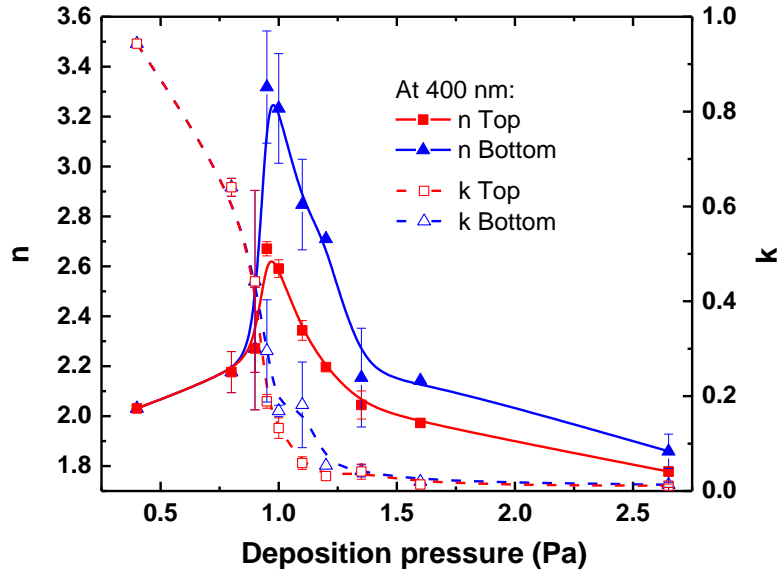

**Fig. 8.** $n$ and $k$ at 400 nm wavelength as a function of sputtering pressure for the $YH_xO_y$ films produced by reactive ($H_2$ to Ar gas flow ratio of 1:2) pulsed-DC magnetron sputtering without intentional substrate heating.

The optical depth profile (at 1.669 eV or 742.8 nm) of the thin films sputtered at different pressures is given in Fig. 9. The previously discovered porosity gradient coincides with the $n$ gradient: the refractive index $n$ decreases in the direction from the substrate to the film's surface as a result of the heterogeneity of the film composition (see next Section 3.5). The lower is the sputtering pressure, the higher is the refractive index difference within the depth of the films; e.g., the films sputtered at 0.90 Pa have -0.25 refractive index difference. The depth profile reduces with an increase of the sputtering pressure reaching a refractive index difference of only -0.07 at 1.35 Pa and 2.65 Pa. A larger optical gradient variation is observed for the first 100 nm of the film thickness, then the refractive index varies very little or remains constant for the rest of the film depth. This has been observed for all films sputtered at pressures higher than 0.95 Pa. There was no optical gradient observed for the films sputtered at pressure < 0.95 Pa.

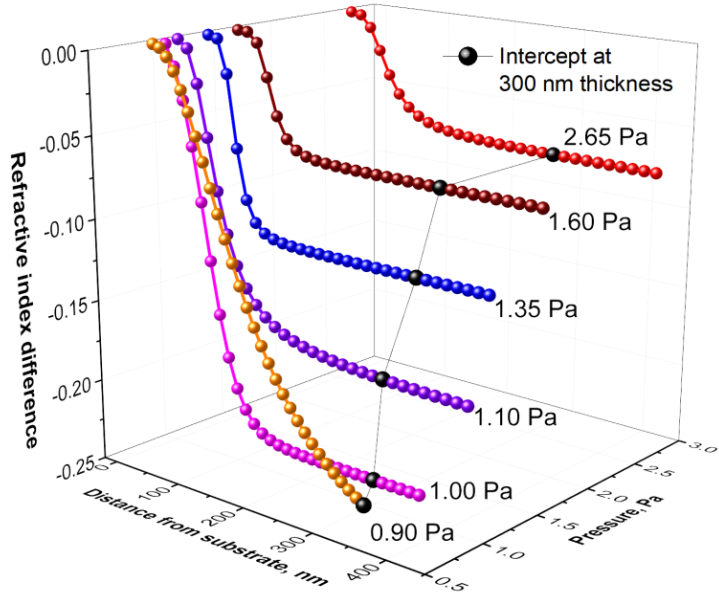

**Fig. 9.** Optical depth profile (at 1.669 eV or 742.8 nm) of the $YH_xO_y$ thin films sputtered at different pressures.

Optical band gap $E_g$ as a function of the sputtering pressure is given in Fig. 10. The values are determined directly as fitting parameters of TLO and HJPS (Herzinger-Johs parameterised semiconductor [41]) functions and from the extrapolation of $\alpha^2(E)$ curves (Tauc plot) obtained from the transmittance data. See the supplementary material for more information on the determination of $E_g$. $E_g$ opens at 0.90 Pa and increases with the deposition pressure. In the case of TLO function, $E_g$ reaches a maximum value of 3.0 eV for the films deposited at 1.10 Pa. With a further increase of sputtering pressure from 1.20 Pa to 2.65 Pa, the values of $E_g$ are slightly lower. This behaviour is somewhat different than what is found in the literature [25], where the band gap increases monotonously from 2.8 to 3.7 eV when the pressure is increased from 0.4 to 6.0 Pa. $E_g$ determined by TLO also gives lower values compared to the HJPS and $\alpha^2(E)$ approaches. First, the procedure used in Ref. [25] is different (using experimental transmittance and reflectance values and assuming a directly allowed transition) from the TLO approach. Assuming an indirect transition in Ref. [34], the values obtained are approximately 0.5 eV lower (in the range of 2.3 to 2.7 eV) and more in line with the TLO results. Second, the thickness (also deposition parameters) of the films in Ref. [25] is different, which influences the determined values of the optical band gap [5,37], most likely due to the grain size. TLO is a very common function in SE to get a fast estimation of an optical band gap even in the case of crystalline materials [42]. This is because TLO uses only four fitting parameters respect to eight in the case of HJPS, thus simplifying the modelling and lowering the correlation between the fitting parameters. In the case of absorption tales (at $E < E_g$) present in the extinction coefficient dispersion curves (like for the YH$_x$O$_y$ films given in the Fig. 7(b)), $E_g$ obtained from TLO are underestimated even for amorphous materials [43,44]. HJPS oscillator function offers precision in $E_g$ evaluation for any material even with absorption tale's issues. The values obtained from HJPS are comparable with the $\alpha^2(E)$ approach. Using $E_g$ and data from Ref. 9, we can conclude that the composition ranges from YH$_{1.25}$O$_{0.87}$ to YH$_{0.56}$O$_{1.22}$ when the sputtering pressure is increased from 0.90 to 2.65 Pa (Fig. 10).

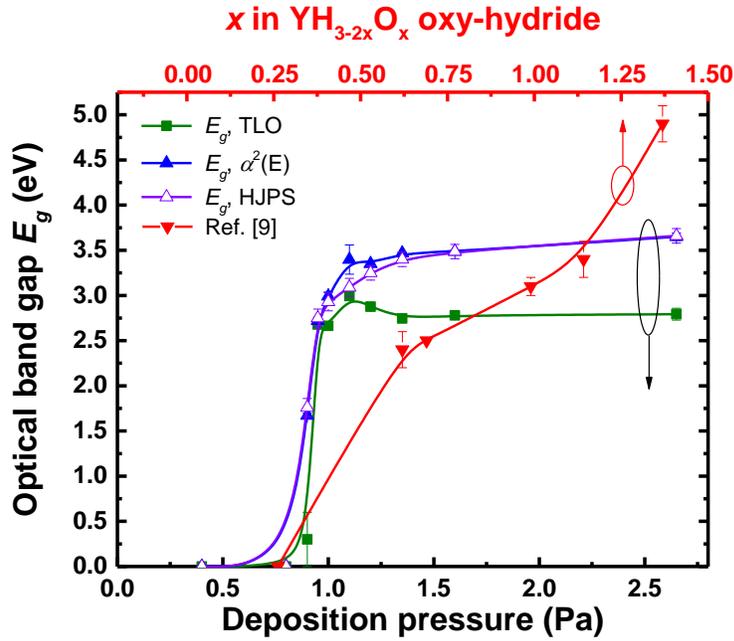

**Fig. 10.** Optical band gap as a function of the sputtering pressure for the YH$_x$O$_y$ films produced by reactive (H$_2$ to Ar gas flow ratio of 1:2) pulsed-DC magnetron sputtering without intentional substrate heating. The values of $E_g$ are obtained directly as a fitting parameter of TLO, HJPS or from extrapolating the linear least squares fit of $\alpha^2$ to zero in $\alpha^2(E)$ plot. The graph also contains $E_g$ as a function of composition (red curve) taken from Ref. 9. Note: the scales of the bottom and top axes are in no way related.

*3.5 Depth profiling by XPS*

The XPS depth profiling measurements were performed to determine the atomic concentration ratio between O and Y throughout the thickness of the YH$_x$O$_y$ films (Fig. 11). For these measurements the films were deposited at the selected pressure values of 0.95 and 1.10 Pa, and on the Ti substrates to reduce electrical charging during the measurements. On the surface, the atomic ratio O/Y is close to the stoichiometry of Y$_2$O$_3$ composition for the both samples. This was expected due to the low electronegativity of Y. A thin surface oxide layer of 5–10 nm has been shown on the non-capped yttrium hydride films by neutron reflectometry [45] and ion beam techniques [8]. The O/Y ratio gradually changes throughout the film thickness indicating the gradient of the composition due to the incomplete oxidation. The values of the O/Y ratio may be slightly overestimated due to the post-etching oxidation in the XPS chamber, which is proved in Ref. [23]. Nevertheless, high O/Y ratios in the range from 0.25 to 0.39 are found, even in the opaque non-capped $YH_xO_y$ films [8]. The reason for the observed optical gradient by SE (Fig. 9) is related to the composition gradient. The changes of O/Y shown in Fig. 11 are steeper close to the substrate, which is in excellent agreement with the SE results. Preferential sputtering of oxygen during the depth profiling cannot be excluded. However, it would be more pronounced close to the surface (rather than the substrate) until steady state is reached, where the surface composition balances the difference in sputter yields. The average O/Y values are 0.80 and 0.84 at the pressure values of 0.95 Pa and 1.10 Pa, respectively. These values are close to those obtained from the optical band gaps in Fig. 10.

The stability of composition for the $YH_xO_y$ film deposited at 0.95 Pa was checked by measuring an XPS depth profile after different storage conditions: immediately after the deposition, after two weeks in the Ar atmosphere, and after two weeks in the air (Fig. S10 in supplementary material). In the first two cases, the curves of the O/Y ratio are almost identical, but in the case of the air exposure, the O/Y ratio has increased. It shows again the continues oxidation over time. However, the gradient of composition is still observed.

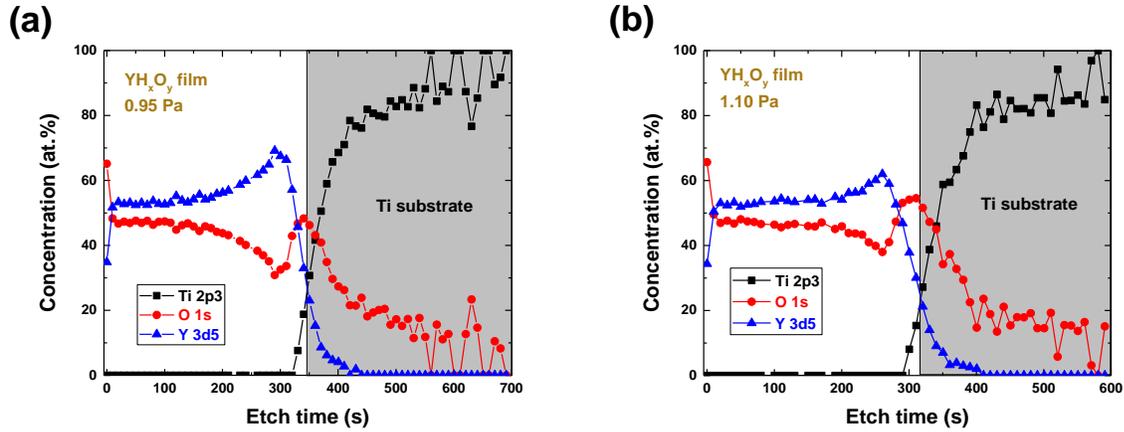

**Fig. 11.** XPS depth profiles of the YH$_x$O$_y$ films deposited at 0.95 Pa (a) and 1.10 Pa (b) by reactive (H$_2$ to Ar gas flow ratio of 1:2) pulsed-DC magnetron sputtering without intentional substrate heating.

## 4. Conclusions

In this paper, the optical properties of the oxygen-containing yttrium hydride films (YH$_x$O$_y$) have been studied *in-situ* and *ex-situ*. The *in-situ* transmittance spectra during the deposition showed that the phase of the films strongly depends on the deposition pressure, even before the exposure to air or pure oxygen. The oxidation due to the residual gas (H$_2$O and O$_2$) levels during the deposition is responsible for this phenomenon. The enhanced oxidation is due to the formation of a more porous structure at higher deposition pressures, confirmed by the electron microscopy images. The time constants of transmittance increase range from a few seconds to two minutes with the tendency to decrease with deposition pressure. The constants reflect the rapidness of the formation of the composition, which ranges approximately from YH$_{1.25}$O$_{0.87}$ to YH$_{0.56}$O$_{1.22}$. The incorporation of oxygen into the films expands the lattice and could lead to the deterioration of the material quality and cause problems in the smart window applications. The chemical composition of these films must be stabilised for long-term use. The YH$_x$O$_y$ films deposited below 1 Pa remain highly absorbing after exposure to air, and $β$-YH$_{2-x}$ is expected to be the main phase. We recommend structure densification and high deposition rates to obtain the purest possible $β$-YH$_{2-x}$ films.

The optical properties *ex-situ* have been studied by spectroscopic ellipsometry. The films can be clearly divided into the two groups: transparent semiconductors/insulators and metallic conductors from the SE analysis. The gradient of refractive index $n$ was found by the optical depth profiles for the transparent YH$_x$O$_y$, and is dependent of the deposition pressure. This is caused by the gradient of porosity and the consequential gradient of composition, which is confirmed by the XPS depth profiles.

**Acknowledgements**

Financial support was provided by Latvian Council of Science Project No. lzp-2020/2-0291. Institute of Solid State Physics, University of Latvia as the Center of Excellence has received funding from the European Union's Horizon 2020 Framework Programme H2020-WIDESPREAD-01-2016- 2017-TeamingPhase2 under grant agreement No. 739508, project CAMART2. The authors thank José Montero and Lars Österlund for the fruitful discussions and ideas on the subject of photochromic yttrium oxyhydride.